# Automated Testing of Graphical Models in Heterogeneous Test Environments

A. Beresnev*, B. Rumpe†, F. Schroven‡

* TU Braunschweig, Software Systems Engineering Institute
† RWTH Aachen, Chair for Software Engineering
‡ Volkswagen AG, Driver Assistance and Integrated Safety

*Abstract*—Automated and therefore repeatable tests are important in product lines to ensure and maintain quality of software functions as well as efficiency of the developers. This article shows methods for fully automated testing of SIMULINK models with the slUnit framework and techniques for the integration of slUnit tests in a general test framework to allow integrated and automated test reports.

## I. INTRODUCTION

Complexity of automotive software systems is still vastly increasing. To achieve the desired optimum of correctness, efficient mechanisms are needed to reproducibly and automatically execute tests and report their findings. Furthermore, integrated testing in heterogeneous development environments is necessary, because we frequently see heterogeneous tool chains including development languages such as C, C++ or Matlab. This article presents improvements to unit test frameworks for SIMULINK [1] and C++[2] which are especially important regarding the automation of test execution and the evaluation of test results.

The article is structured as follows: The first section describes the technical basis of automated software testing and graphical models used in the automotive industry. Then a test automation system for C++ is presented along with slUnit [1] – a unit test environment for SIMULINK. Finally two methods are shown to execute slUnit tests fully automatically.

## II. TECHNICAL BASIS

### A. Automated Software Tests

Since the idea was made prominent by Kent Beck [3] unit tests have become an essential part of the quality assurance in modern software engineering. Not only do unit tests help to find software problems early in the development cycle where the cost of fixing bugs is minimal. Fully automated tests moreover are good indicators that the desired functionality has been stable after any change e. g. necessary as enhancement of the requirements. The whole development process becomes more flexible and predictable and thus is more amenable for product lines [4]. That is why unit tests are vital for agile software development techniques like eXtreme Programming (XP) [5], [6] and Test Driven Development (TDD) [7] which imply the creation of unit tests even before the actual functionality is implemented.

One of the main principles of unit tests is that the test specification language is the programming language itself, i. e. the language of the code under test. In fact this was one of the main reasons of unit tests' success: the developers can use their familiar programming language and do not need to learn and switch to another just to write tests.

We now do have unit test frameworks (e. g. JUnit, CppUnit, SlUnit) for nearly every existing programming language. Their main goal is to make the development, execution and reporting of unit tests as comfortable as possible. Here, the programmer should just immediately and efficiently be able to write the code of a test method including automated check of the result, push a button and see the tests running ok (green) or failure (red). For this reason xUnit frameworks often facilitate integration directly into an IDE or build system.

Since unit tests run automatically at no cost they can be repeated after every change in the software, along with incremental or at least daily builds of the entire project. This technique is primarily known under the term "Continuous Integration" and is basically independent of other XP principles. Under the control of a continuous integration system, unit tests are most effective, since every bug which can be detected by existing unit tests is almost immediately reported after it is created and the developer should be able to find and remove it much easier. From our own experience (e. g. [8]) we know that a continuously integrated project is much more predictable, since one always knows which parts of the system already work correctly, which still do not, and which bugs are to be fixed next. In later development phases and evolutionary follow-up projects, continuous integration makes it possible to always have the last fully functional version of the software which can be deployed immediately.

## B. Graphical Models for Automotive Engineering

In automotive engineering, model based design and automatic code generation have reached significant importance. In this context, graphical models, for example MATLAB/SIMULINK [9] or UML [10], [11], are used for coding high-level artifacts.

One primary goal of those models is to generate code that is compiled for a certain target. In the context of automotive research projects, dSPACE rapid prototyping hardware [12] is quite often used. The generator supports common vehicle buses (e. g. CAN) on the level of signals and thus allows fast integration in demonstrator vehicles. Together with online parameter modification and measurement of values this speeds up development and manual testing.

The model can also be used for offline simulation. Then all interfaces to the car need to be simulated as well, but the simulation can run efficiently overnight and without any physical device (car) attached. In case of embedded automotive software development this means simulating directly connected switches and data from other electronic control units (ECU) that would be received over an appropriate bus. Two cases can be distinguished:

- Offline simulation
- Interactive simulation

In case of *offline simulation* generated or recorded input data is used to feed the control system. The systems output is usually saved in files or evaluated directly.

Although vehicle integration is quite comfortable, in the long run testing in a real car is far more time consuming than simulating. It is also quite easy to build an *interactive simulator* around a SIMULINK model. Therefore, manageable situations can easily be investigated using an interactive simulator. Depending on the type of application, a more or less explicit and detailed model of the vehicle and its surrounding environment is needed. In the intercative case the output, i. e. the car's behavior, is usually directly visualized.

In both cases it is necessary to set up the model so that it automatically detects whether it is used in simulation or for code generation to dynamically switch on and off parts of the model that are only conditionally needed.

## III. UNIT TESTING WITH CXXTEST AND SLUNIT

Existing unit test frameworks, namely CxxTest [2] for C++ and slUnit [1] for Simulink can be used for two very different development environments. However, both need to be adapted and in particular integrated to allow heterogeneous developments in C++ and Simulink to be tested in an integrated fashion.

### A. Continuous Integration with CxxTest

Unit test frameworks for Java, .NET and other languages heavily rely on reflection mechanisms to automatically identify the methods and classes containing the test code and thereby minimize the amount of code required to define a test case or test suite. Due to the lack of reflective facilities in C++, unit test frameworks for this language (like for example CppUnit) oblige the developer to mark test cases explicitly through macros or function calls which however contradicts the spirit of xUnit. CXXTEST however uses the technique of parsing the source code before compilation and generating a test runner which is able to call the test functions directly. This approach allows the test suites to be as simple as shown in figure 1. The only requirements are that a header file is created with a class derived from `CxxTest::TestSuite` and the test methods' names must start with `test`.

```
// MyTestSuite.h
#include <cxxtest/TestSuite.h>

class MyTestSuite : public CxxTest::TestSuite
{
public:
    void testAddition( void )
    {
        TS_ASSERT( 1 + 1 > 1 );
        TS_ASSERT_EQUALS( 1 + 1, 2 );
    }
};
```

Fig. 1. A simple CxxTest example

To generate the test runner for one or several such test suites the developer would run the generator for example as follows:

```
cxxtestgen.pl --error-printer -o \
runner.cpp MyTestSuite.h
```

This would generate `runner.cpp`, which then could be compiled as usual.

Based on CxxTest, we have implemented BUG-BUSTER, a Continuous Integration System. BUGBUSTER is able to run on any machine which can access the source repository of the project. It checks the repository periodically for changes and, if an unchecked revision is detected, initiates a configurable action sequence. Following actions are available:

- Check out
- Build
- Run unit tests
- Measure code coverage
- Generate reports for the web
- Optionally mail report result to developers
- Clean up

BUGBUSTER contains a test coverage tool indicating the quality of the tests in terms of coverage (statements and control flow on code level, but not yet on diagram level). As usual, it instruments the C++ code by inserting extra code around each instruction. E. g. the instruction's line number is passed as a parameter. During the test run, the coverage measurement tool matches the line numbers with the corresponding source code and has the ability to determine the percentage of the instructions actually being executed.

To generate HTML reports like in figure 2, BugBuster enhances CXXTEST with a test runner which produces XML-based output of the test results instead of presenting them in a console or a GUI window. Subsequently, the test outputs are processed by a set of XSLT transformations, which combine them with the results of the coverage measurement and the extracted code fragments to a comprehensive test report which not only shows the number and status of the executed tests but also gives an impression of their quality. The extracted code fragments allow the developers to quickly look up the position and cause of a failed test directly in the HTML report. Finally, BUGBUSTER can send the compiled test reports via email to the developers or store them in a shared folder.

An important feature of BUGBUSTER is its revision management which traces not only the version of the main project repository but also all of its external dependencies. For this purpose BUGBUSTER, maintains its own list of virtual revisions. Each of them is composed of the main project's revision number and the revision numbers of all externals. Every change of these numbers causes BUGBUSTER to create a new virtual revision. This allows the test results of the project to be monitored separately with every version of all subprojects and external libraries.

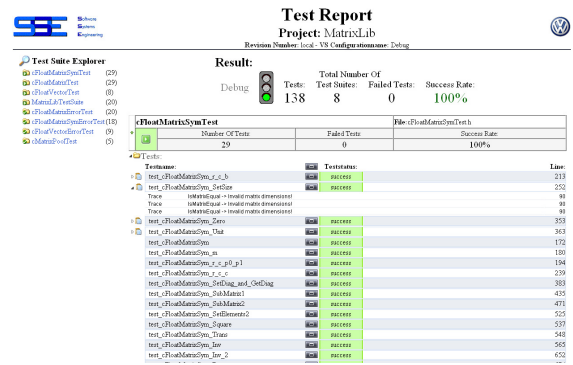

Fig. 2. BugBuster HTML report

*B. Interactive slUnit*

SlUnit is an interactive testing framework for SIMULINK models [1]. It is completely implemented in the language of MATLAB/SIMULINK and therefore follows one of the basic ideas of xUnit testing, i. e. to implement the test framework in the same language as the components to be tested. To construct a testbed, one can use templates that already contain an empty test case, a placeholder for the system under test (SUT), the GUI elements to control the test execution and a progress bar showing the testsuites execution state. It is possible to run a single test case or all tests of the test suite. The result of each test case is displayed through the color of the corresponding subsystem. Such a setup allows users to develop their model and interactively see whether the changed systems still fulfills the requirements that are represented by the tests.

slUnit uses the following test patterns: *assertions*, *test methods*, *all tests* and *fixtures*. The *assertion* is the basic element for testing that compares the systems output for a given input with an expected value. In case of equality, the test passes, otherwise it fails. When an error occurs when executing a test, the test is aborted and slUnit proceeds to the next. This ensures that tests are independent from each other, which is another of the xUnit principles [13]. As SIMULINK is made for dynamic systems testing means simulating the model for a certain period of time. If all signals and transfer functions within the model are constant with respect to time, it is sufficient to minimize the simulation to speed up testing.[1] The test methods in-clude processing blocks for the test case. At least the input and expected output values are given. Fixtures share common code that is needed for all test cases but is not part of the system under test. This can easily be implemented in SIMULINK and is just one block that is located "beyond" the multiplexer and hence is used for all test cases. This block could for instance be used for input conversion that is the same for all test cases.

---
[1] It is sufficient that the assertion fails in a single time step.

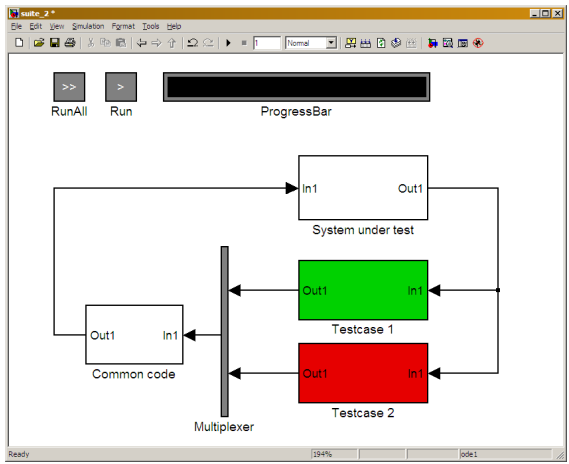

Fig. 3. slUnit testsuite with two testcases, shared code, the system under test and GUI elements

## IV. AUTOMATION AND INTEGRATION OF TESTING FRAMEWORKS

### A. Fully Automated slUnit

One of the most important principles in unit testing, *automation* of the test run and the result examination, has so far only been realized to a small extent, i. e. that all tests in a test suite can be run consecutively while the SIMULINK model is open. When working on a larger project, there will be many modules to test repeatedly. Therefore, the need for automated processing is evident.

slUnit had an intuitive mechanism to show the test results: the test cases subsystem block is colored green in case of all assertions passing and red when at least one assertion fails (figure 3). The colors are saved with the model, so that one can always see the results of the latest run. When tests are spread over several suites, i. e. several files, it, however, becomes quite impossible to get an overview over the project's status.

In our approach both shortcomings have been overcome. Concerning automation, slUnit can be controlled from a *test runner*, which only needs the specification of a directory and the names of the test suites to be included. Figure 5 shows an exemplary script.

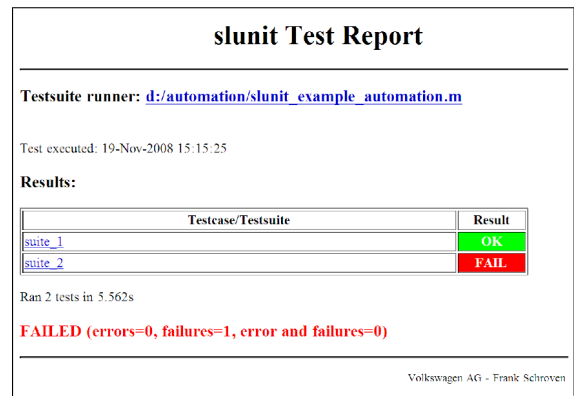

Fig. 4. slUnit overview HTML report of two small testsuites with hyperlinks to the detailed test results

After setting a root directory all test suites' names are aggregated in a cell string[2]. After that, the test runner is called with two more arguments: the function `mfilename` returns the file name of the calling script which is used to generate a plausible file name for the report file so that the user can easily associate the test script and the HTML report. Besides that, the argument *verbosity* (here 1) describing the extent of the report is passed to the test runner.

```
testpath   = '<directoryString>';
testsuites = {'suite_1',...
              'suite_2',...
             };

slunit_testrunner(testpath, testsuites,...
                  mfilename, 1);
```

Fig. 5. Example of a slUnit testrunner call

The test runner ensures that all test suites are executed and takes care of the result files. The test runner can easily be wrapped in a batch file so that execution (including startup of MATLAB) can be planned without any user interaction and thus fully automated. The test runner avoids dependencies among the test suites, similar to the single tests. It means that if one test or a whole test suite exhibits problems, such as a crash because of an error, there is no effect on the execution of the others.

The test results are presented in a hierarchical report that is built by a custom HTML generator. The report is interactively examinable at different levels of detail. A call of the test runner leads to *one* summary that sums up the results for all test suites (Figure 4) and gives the project manager quick feedback and the possibility to take actions when necessary. The re-port also informs about the time and duration of the test execution. For further investigation, the summary can contain hyperlinks to the result files for every test suite. These low level reports contain hyperlinks to the executed test suites (*model files*). That means that there is a continuous path from a high level summary down to the test case of interest.

---
[2]A cell string is a MATLAB data type that can handle strings of different lengths.

The productive part, the system under test, can either be developed within the test suite or not. When working alone on a small project developing within a set of test cases might be possible. As a project is larger this is impossible because models grow and will most likely be divided into smaller modules that are tested individually. Since a test suite is an ordinary SIMULINK model the developer has to ensure that no inconsistency arises from the fact that test suite and the developed model are different files. This can be achieved by using the techniques of *libraries* or *model referencing* (figure 6).

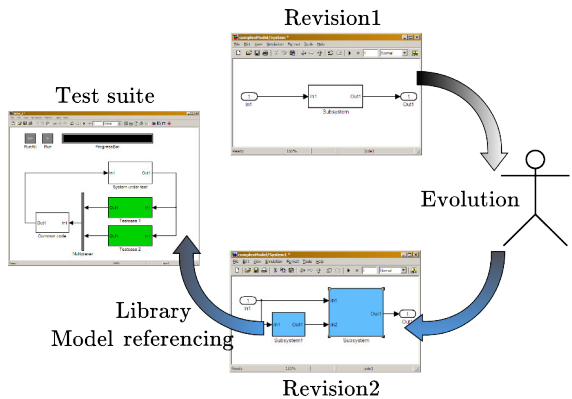

Fig. 6. Automatic update of slUnit test suites when working with libraries or model referencing

### B. Integrating Heterogeneous Test Frameworks

Despite the growing importance of model-based development and code generation in the automotive industry, projects usually can not be implemented using only MATLAB/SIMULINK and would be to low-level to only use C++. So a mixture of languages is used. This leads to heterogeneous development and test environments and if tests are more than just method or module tests, both tested system and tests are heterogeneous. Our experience e.g. in the CAROLO project [14] shows how important it is to integrate and test the modules from such different worlds early. A good integration also allows the creation of test parts which interact with modules from another. Another positive side effect of this solution is the consistence of test reports, i.e. the status of the entire project is presented in a single report.

In some projects, we integrate slUnit test cases into a BUGBUSTER test environment which already existed for the C++ part of the project [15], [16]. However, as mentioned in section III, BUGBUSTER expects to find test suites in CXXTEST style and can not handle SIMULINK models of slUnit directly. Therefore, we embed slUnit test cases to make them look like CXXTEST test cases. More precisely, we use adapters for each slUnit test suite which can be executed and evaluated by the CXXTEST framework. We use a generator, developed for that purpose, which scans all SIMULINK models in a given directory producing an adapter for each detected slUnit test case of the form shown in figure 7 into a prepared CXXTEST test suite.

```
void test_<<TestCaseName>>()
{
    << run slunit test >>
    << get result and output from matlab >>
    << forward output >>
    if( << result is negative >> )
    {
        << fail CxxTest >>
    }
}
```

Fig. 7. slUnit test case adapter for CxxTest

To actually run a slUnit test case from C code and obtain the necessary information like test results we utilize the external interface of MATLAB [17] which allows to call MATLAB software from other programs, thereby employing MATLAB as a mathematical library or as a simulink engine. During the execution of our adapted slUnit tests the MATLAB engine is initialized only once prior to all tests as a global test fixture. After that every test starts a MATLAB script passing the test case name as parameter. This script executes a single slUnit test case end returns its results.

Of course executing slUnit test cases in context of an existing C++ unit test environment can be seen as a special case of a more general problem, namely the integrated execution of test cases for two or more different unit test frameworks with the aim of using a single continuous integration system or reporting mechanism. We have positive experiences with the described approach of generating adapter test cases.

### V. CONCLUSIONS AND OUTLOOK

This article presents an adaptation of the SIMULINK testing framework slUnit and its integration into C++-based tests. Two ways for a full automation including a convenient reporting mechanism were discussed.

Of course, we can improve our tool, e. g. providing more information than labeling the test results as *passed*, *failed* or *error*. In case of *failed* further information, e. g. about internal values and their time dependent course need to be added to the report. Future evolutions of slUnit will therefore include data sinks that store values to plot them into the detailed reports.

However, the most important task for now is to actively help traditional developers to adopt this testing technique and use it for early testing as well as for integration tests. This is a continuously hard job, as it strongly affects the way how implementation is carried out. Interestingly, developers in business projects have successfully adopted these mechanisms of the recent years and we support the hypothesis, that embedded developers will also be able to use this kind of testing techniques and improve efficiency and if desired quality of their products. Of course, in the embedded systems area this is somewhat more complicated, as traditional real-life tests of software and controlled device always will be necessary in addition to pure software tests – but hopefully considerably less.

## References

[1] Thomas Dohmke and Henrik Gollee, "Test-Driven Development of a PID Controller," *IEEE Software*, vol. 24, no. 3, pp. 44–50, May 2007.

[2] Tigris, "CxxTest," http://cxxtest.tigris.org, 2008.

[3] Kent Beck, "Simple smalltalk testing: With patterns," 1989.

[4] Klaus Pohl, Günter Böckle, and Frank van der Linden, *Software Product Line Engineering. Foundations, Principles, and Techniques.*, Springer, Berlin, 2005.

[5] Kent Beck, *Extreme Programming Explained: Embrace Change*, Addison-Wesley, 1999.

[6] Bernhard Rumpe, "Extreme Programming Back to Basics," *Modellierung*, pp. 121–131, 2001.

[7] Kent Beck, *Test Driven Development. By Example*, Addison-Wesley Longman, 2002.

[8] C. Basarke, C. Berger, K. Berger, K. Cornelsen, M. Doering, J. Effertz, T. Form, T. Gülke, F. Graefe, P. Hecker, K. Homeier, F. Klose, C. Lipski, M. Magnor, J. Morgenroth, T. Nothdurft, S. Ohl, F. Rauskolb, B. Rumpe, W. Schumacher, J. Wille, and L. Wolf, "Team CarOLO – Technical Paper," Tech. Rep., Technische Universität Braunschweig, Carl-Friedrich-Gauss-Fakultät, 2008.

[9] The Mathworks, "Matlab & Simulink," http://www.mathworks.com, 2008.

[10] Bernhard Rumpe, *Agile Modellierung mit UML*, Springer, Berlin, 2004.

[11] "Unified Modeling Language Specification," http://www.omg.org/spec/UML/, OMG, 2008.

[12] "dSPACE," http://www.dspace.com, 2008.

[13] Gerard Meszaros, *xUnit Test Patterns: Refactoring Test Code*, Addison-Wesley Longman, 2007.

[14] Christian Basarke, Christian Berger, and Bernhard Rumpe, "Software & Systems Engineering Process and Tools for the Development of Autonomous Driving Intelligence," *Journal of Aerospace Computing, Information, and Communication*, vol. 4, no. 12, pp. 1158–1174, October 2007.

[15] Andreas Weiser, Arne Bartels, and Simon Steinmeyer, "Intelligent Car: Teilautomatisches Fahren auf der Autobahn," in *Tagungsband der AAET 2009 – Automatisierungssysteme, Assistenzsysteme und eingebettete Systeme für Transportmittel*, 2009.

[16] Christian Berger, Holger Krahn, Bernhard Rumpe, and Arne Bartels, "Qualitätsgesicherte Fahrentscheidungsunterstützung für automatisches Fahren auf Schnellstraßen und Autobahnen," in *Tagungsband der AAET 2009 – Automatisierungssysteme, Assistenzsysteme und eingebettete Systeme für Transportmittel*, 2009.

[17] "Calling MATLAB Software from C and Fortran Programs," http://www.mathworks.com/access/helpdesk/help/techdoc/matlab_external/f38569.html.